\documentclass[aps,prl,reprint,twocolumn]{revtex4}
\usepackage{graphicx}% Include figure files
\usepackage{epsfig,subfigure,float,floatflt,wrapfig,float}
\usepackage[bbox,pdftopdf=true,suffix=]{epspdfconversion}
\usepackage[english,]{babel}
\usepackage{enumitem}
\usepackage{siunitx}
\usepackage{soul,xcolor}
\usepackage{dcolumn}
\usepackage{bm}
\begin{document}
\title{Unprecedented confinement time of electron plasmas with a purely toroidal magnetic field in SMARTEX-C}
\author{Lavkesh Lachhvani}
\email{lavkesh@ipr.res.in}
\affiliation{Institute for Plasma Research, HBNI, Gandhinagar-382428,India}
\author{Sambaran Pahari}
\altaffiliation{Bhabha Atomic Research Centre, HBNI, Visakhapatnam-530012, India}%
\author{Rajiv Goswami}
\affiliation{Institute for Plasma Research, HBNI, Gandhinagar-382428,India}%
\author{Yogesh G. Yeole}
\affiliation{Institute for Plasma Research, HBNI, Gandhinagar-382428,India}%
\author{Prabal K. Chattopadhyay}
\affiliation{Institute for Plasma Research, HBNI, Gandhinagar-382428,India}
\date{\today}
\begin{abstract}
Confinement time of electron plasmas trapped using a purely toroidal magnetic field has been extended to $\sim 100$ s in a small aspect ratio ($R_{o}/a \sim \num{1.59}$, $R_o$ and $a$ are device major and minor radius, respectively), partial torus. It surpasses the previous record by nearly two orders of magnitude. Lifetime is estimated from the frequency scaling of the linear diocotron mode launched from sections of the wall, that are also used for mode diagnostics. Confinement improves enormously with reduction in neutral pressure in the presence of a steady state magnetic field. In addition, confinement is seen to be independent of the magnetic field, a distinguishing feature of Magnetic Pumping Transport (MPT) theory. Since MPT predicts an upper limit to confinement comparisons have been made between our experiments and MPT estimates.
\end{abstract}

\maketitle

%Introduction-1 (a): Application of Toroidal NNP- Of general interest to Physics Community
The excellent confinement of electron plasmas in cylindrical traps had unleashed a plethora of laboratory investigations into their rich 
collective dynamics in the latter half of previous century \cite{Malmberg_75_PRL,driscoll_thermal_EQ_1988,DubinONeil_RevModPhys_1999}. This impacted a large number of fundamental studies relevant to diverse fields ranging from atomic physics to incompressible fluid dynamics. In contrast, the behaviour and applications of such single species plasmas confined in other geometries and magnetic field topologies remained rather unexplored due to their unknown confinement properties. Historically though, experiments in toroidal electron plasmas, preceded cylindrical plasmas and several applications had also been proposed. These included, for example, formation of a deep potential well by an electron cloud trapped in toroidal geometry as a source of highly stripped ions \cite{Janes_HIPAC_PR_145_925_1966}, heavy ion particle accelerator \cite{Daugherty_Hevayions_PRL_20_369}, electrostatic thermonuclear fusion reactor \cite{Fusion_NNP_PRL_Turner_70_798_1993,Fusion_Stix_PRL_24_135_1970}, or as a shielding mechanism for space vehicles from solar radiation 
\cite{french_plasma_RadiationShield_1968}. Besides these, confinement of non-neutral plasma in toroidal geometry and investigating the effects of arbitrary degree of non-neutrality under controlled conditions~\cite{sarasola_first_2012} was also expected to aid the understanding of transport in neutral plasmas which are of profound interest to the fusion community. In recent times, much of the motivation and interest in toroidal traps seem to follow from the possibility of creating electron-positron pair plasmas 
\cite{Surko_EP_PRL_75_3846,Pedersen_E-P_Plasma_NJP_2012} due to the expected lack of instabilities in such plasmas and in view of their 
relevance to astrophysical objects. In addition to this, just like cylindrical electron plasmas in homogeneous magnetic field have served as excellent test-beds for carrying out incompressible fluid dynamics experiments~\cite{Fine_SymmVortexMerger_PRL_67_588,Fajan_VortexinVortex_PRL_85_4052}, toroidal electron plasma in the presence of an in-homogeneous magnetic field may mimic compressible fluids and has remained an attractive proposition for some time \cite{Ganesh_coherent_Frontiers_Turbulence_2006} that certainly merits further investigation. 

%Introduction-1 (b): Confinement limitation in toroidal traps -- Confinement achieved so far & MPT background
Undoubtedly, the minimum underlying requirement for achieving any of the stated objectives with toroidal non-neutral plasmas, is a long time confinement. Theoretically, in the presence of a purely toroidal B field, such plasmas are supposed to be in stable equilibrium 
\cite{daugherty_equilibrium_1967,Avinash_Equilibrium}). However, unlike cylindrical plasmas in uniform magnetic field which are governed by robust confinement theorem~\cite{Oneil_Conf_Theorem_PoF_1980}, plasmas trapped with a purely toroidal B field are thought to be fundamentally limited in their confinement properties due to MPT. Proposed by O'Neil and Crooks 
\cite{ONeil_Crooks_MagneticPumping_PoP1996}, this radial transport arises due to $E \times B$ drifts of the plasma in a spatially 
inhomogeneous toroidal B field. The flux rates, as derived in the limit of large aspect ratio, suggest a transport time scale
\cite{stoneking_experimental_2009} that is dependent on major radius $R_{0}$ and electron temperature $T_{e}$. Experimentally, a few early initiatives in toroidal traps reported successful trapping to varying degree and successfully demonstrated steady state confinement of a few hundred microseconds \cite{daugherty_experiments_1969,Clark_Expt_ToroidalInj_PRL_37_592,zaveri_low-aspect-ratio_1992,khirwadkar_steady_1993} overcoming the single particle drifts. Renewed interest in toroidal traps in the late 90's and early 2000s, led to a major turnaround. Two of the new traps with purely toroidal magnetic field, were converted into partial torus in order to combine the technique and advantages of cylindrical traps with toroidal geometry. Among them were LNT-I~\cite{stoneking_electron_2002} which was a large aspect ratio device and SMARTEX-C \cite{pahari_electron_PoP2006}, a small aspect ratio trap. While significant improvement in trapping ranging from few to 10s of ms were achieved, yet, till early 2006, the best confinement times reported remained orders of magnitude less than those predicted by MPT theory. Finally, in 2009, with improved operating scenarios like enhanced vacuum, higher magnetic field and a higher degree of symmetry LNT-II successfully confined toroidal electron plasma in a steady state for $\SI{3}{\second}$ \cite{Stoneking_1S_confinement_PRL2008,stoneking_experimental_2009,ha_using_2009}. They argued that the confinement time approached the limit set by MPT (for the trap major radius of $\SI{17.4}{\centi\meter}$ and assumed temperature of $\SI{1}{\electronvolt}$), although no direct evidence of MPT was observed. Results from SMARTEX-C followed, whose reported confinement time ($\num{2.14} \pm \SI{0.1}{\second}$) was also very close to their theoretically predicted time scales \cite{lavkesh_PoP_confinement_1S_2016}. Its worth mentioning here that no such theoretical limitations are known to apply to plasmas trapped in other B field topologies like in Stellarators or, say, in a torus with levitated dipole magnetic field. Attempts to confine such plasmas on nested flux surfaces in a Stellarator \cite{CNT_Pedersen_90ms_PoP_2012} though has been limited to $\sim$ $\SI{90}{\milli\second}$ due to ion dynamics. Experiments with a levitated dipole \cite{RT_Levitated_Dipole_PRL_104_2010} magnet have however triumphed in extending the confinement to $\sim$ $\SI{300}{\second}$ and has been by far the most promising. It may thus appear that low temperature electron plasmas trapped {\it with purely toroidal B field} are fundamentally constrained and will perhaps never be able to achieve the goals of long time confinement and thermal equilibrium like their cylindrical counterparts and/or other contemporary traps with alternate B field topologies.   
 
This letter reports an improvement in the confinement of pure electron plasma by nearly two orders of magnitude in an upgraded 
SMARTEX-C, where plasma remains in a steady state equilibrium for nearly $\num{100} \pm \SI{20}{\second}$ at $200$ Gauss. This is the highest reported time from a trap with purely toroidal B field. Comparisons of experimental results with predictions based on MPT are interesting and have been revisited.
%\textbf{\textit{Introduction-1(c): Put SMARTEX-C Trap geometry and details}}
\begin{figure}[ht!]
\begin{center}
\includegraphics[width=0.5\textwidth]{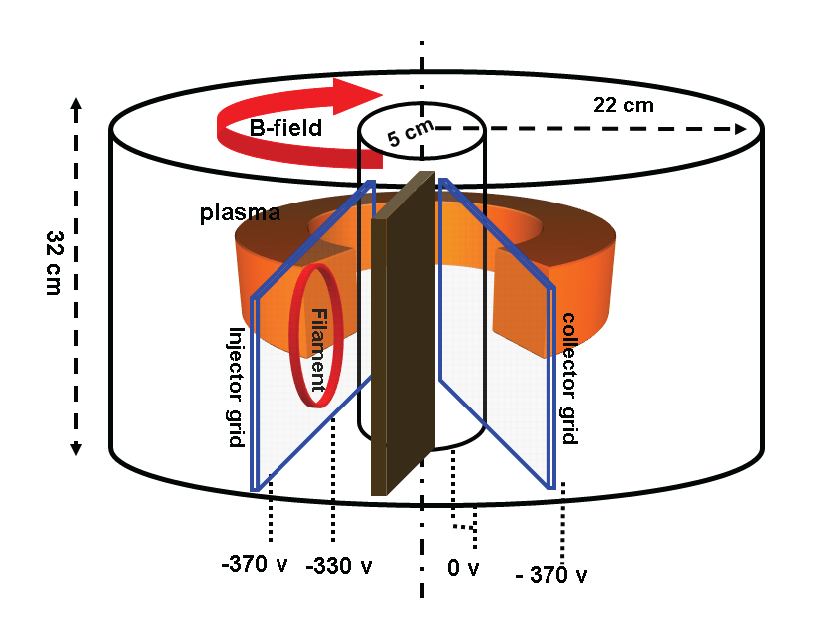}
\caption{Schematic diagram of SMARTEX-C device}
\label{Figure:Schematic_SMARTEXC_Device}
\end{center}
\end{figure} 

SMARTEX-C is a partial (C-shaped) toroidal trap with aspect ratio of $R_{0}/{a} \sim 1.59$, trapping electron plasma in an angular arc of $\Phi \sim 315^o$. The electrode arrangement as shown in Fig. \ref{Figure:Schematic_SMARTEXC_Device} allows us to operate the trap in a ``inject-hold-dump'' cycle~\cite{pahari_electron_PoP2006}, as is typically carried out in cylindrical Penning-Malmberg traps.  
The electrons emitted thermionically are injected for a brief period ($\sim \SI{60}{\mu\second}$) into the arc by turning the injector 
grid bias positive with respect to the (tungsten) filament. As the grid turns negative, the injection stops. The electrons injected at $\sim \SI{250}{\electronvolt}$ parallel energy are held between negatively biased end-electrodes and radially confined with a purely toroidal B field. A sufficient number of these electrons gives them a collective $E\times B$ drift that helps to overcome the single particle drifts, thus forming a plasma. It may be noted that presence of neutrals contaminates a pure electron plasma primarily affecting its confinement. In our earlier experiments pressures of $\sim \SI{4e-9}{\milli\bar}$ were reported from a gauge placed close to the pumping port. As the confinement time has been observed to be highly sensitive to neutrals, pressure measurements were re-calibrated accurately with a nude Bayard-Alpert gauge located close to the trapping arc; the measured pressure in such cases was found to be higher by a factor $\sim \num{4}$ ($\num{1.5} \pm \SI{0.1e-8}{\milli\bar}$). To improve the vacuum a Non-Evaporable Getter (NEG) pump was installed on a port close to the trapping arc. Background pressures in the trapping arc now improved to $\num{5.0} \pm \SI{1.0e-9}{\milli\bar}$. In earlier experiments thermal loading of toroidal field (TF) coils (AWG-8, 20 turns) limited the steady state B-field ($\sim 380$ Gauss) to $4$ s with a droop of $\sim 35\%$. The confinement time was found to be sensitive to the droop rather than the strength of the magnetic field. The limited operation to $4$ s also made any estimation of long time confinement unreliable. Upgraded TF is now generated out of an AWG-2 silver-plated, 24 turns copper cable. A maximum of $900$ Gauss could therefore be generated for $\num{3}-\SI{4}{\second}$ with $< \num{5}\%$ droop. Whereas, a nearly steady state B field for $\sim 40$ s could be generated at $\num{100}-\num{200}$ Gauss with droop $< \num{0.1}\%$. The latter has given us the best and most reliable estimates of confinement time in SMARTEX-C. 
Principal diagnostics of the trap are capacitive probes~\cite{kapetanakos_diagnostics_1971} which are essentially parts of wall insulated 
from the rest of it and are located at various toroidal and poloidal locations. These wall probes are utilized to monitor image currents that can be interpreted to obtain information about any electrostatic activity in the plasma~\cite{Lavkesh_Dioc_PoP2017}. Additionally, if the plasma is quiescent, these probes have been used to excite normal modes, namely the diocotron modes that are ubiquitously present in such single species plasma. Under linear approximations, the frequency of $m=1$ mode, which represents the azimuthal rotation of a displaced charge cloud, is a good estimate of the charge content. 
\begin{figure}[ht!]
\begin{center}
\includegraphics[width=0.5\textwidth]{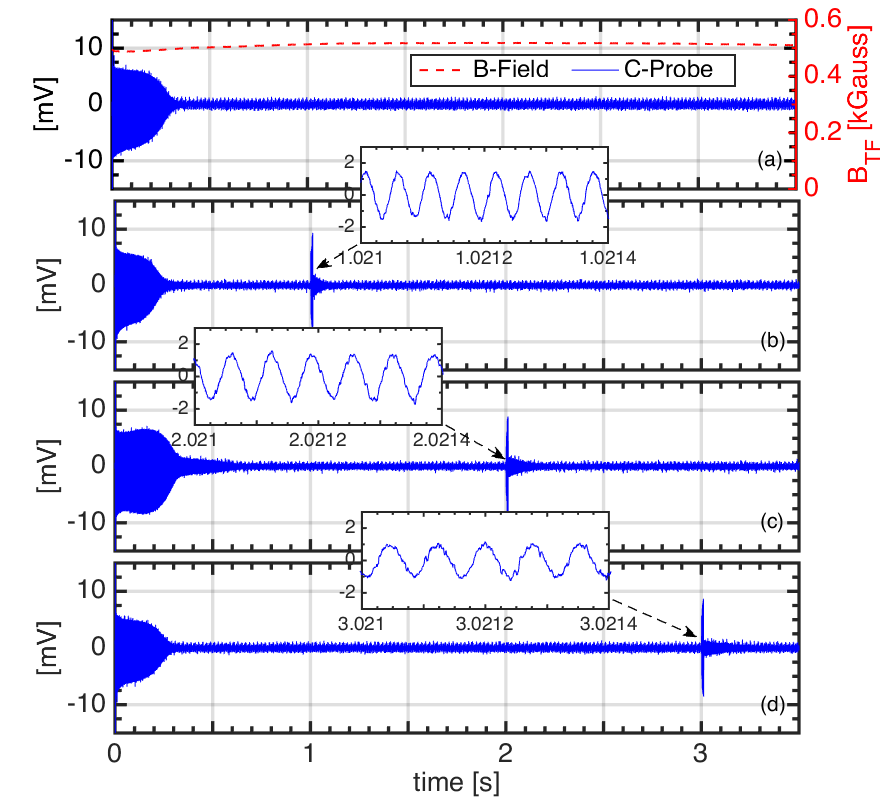}
\caption{Time trace of capacitive probe oscillations (a) before launching and (b) after launching diocotron mode externally at (b) 
$\SI{1.02}{\second}$, (c) $\SI{2.02}{\second}$, and (d) $\SI{3.02}{\second}$ for magnetic field of $\num{525}$ Gauss.}
\label{Figure:Launch_Dioc_diffTime_500Gauss}
\end{center}
\end{figure} 
In cylindrical machines, such modes have been therefore used as a non-destructive diagnostics to estimate the total stored charge. In toroidal machines, if the mode is linear, the total charge content $Q$ can be obtained from mode frequency ($m=1$) using $f_{D} = Q/4\pi^2 \epsilon_{0} L_{p} R_{w}^{2} B$, where $Q$ is total stored charge, $L_{p}$ plasma length, $R_{w}$ wall radius and B is toroidal magnetic field. The frequency evolution can therefore provide us with an estimation of confinement time ~\cite{lavkesh_PoP_confinement_1S_2016}. Note that in toroidal electron plasmas, shift in equilibrium position, if any, has to be accounted for and B field at equilibrium position has to be used. However estimation of lifetime from the time evolution of frequencies is unaffected by this shift as it only scales the exponential function by a factor.   

\begin{figure}[ht!]
\begin{center}
\includegraphics[width=0.5\textwidth]{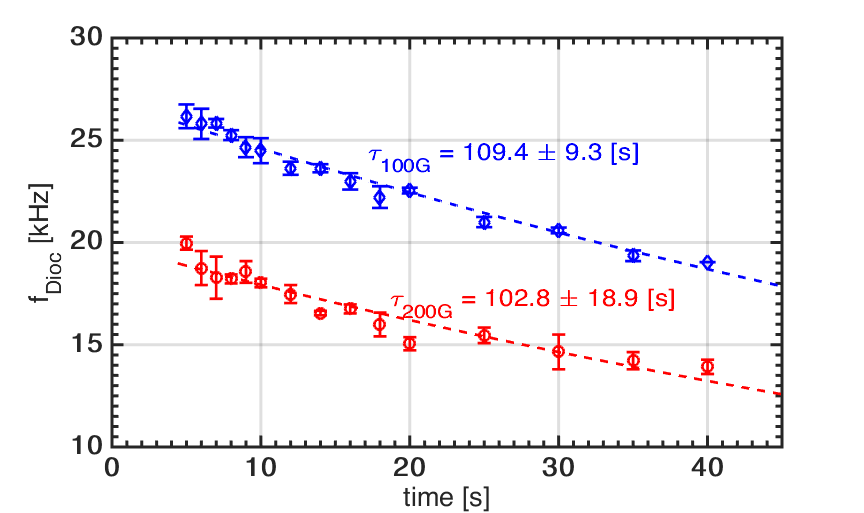}
\caption{Observed diocotron mode frequency on capacitive probe along with exponential fit for low magnetic fields of $100$ and $200$ Gauss, at low pressure of $5.0 \times 10^{-9} \pm 1.0 \times 10^{-9}$ mbar and injection energy $V_{Inj} = 250$ V.}
\label{Figure:Tau_100_200Gauss_100S}
\end{center}
\end{figure}

In a typical experiment, a natural diocotron mode appears soon after injection but quickly damps in less than one-fourth of a second resulting in a quiescent plasma (See Fig ~\ref{Figure:Launch_Dioc_diffTime_500Gauss}(a)). Such a quiescent plasma in stable equilibrium has been achieved through a series of measures leading to control of various instabilities~\cite{lavkesh_PoP_confinement_1S_2016}. With instabilities arrested, charge loss rate reduces and lifetime increases. However, in such a quiescent plasma with no electrostatic activity, wall probes fail to detect any image current and signatures of plasma dynamics. In such a scenario the presence of plasma can be detected by launching the diocotron mode externally. A small electric field perturbation is applied to the equilibrium plasma using a pair of oppositely located wall probes at a toroidal location ~\cite{lavkesh_PoP_confinement_1S_2016}. While the applied perturbation may contain a chirp of frequency within a few kHz, the linear $m=1$ mode is excited with a frequency that is proportional to the charge content in the plasma at the instant of launch. The launched mode at $\num{5e2}$ Gauss at different instants after injection are shown in Figure ~\ref{Figure:Launch_Dioc_diffTime_500Gauss} (b)-(d). Several ``injection-hold-launch'' cycles have been repeated keeping all other operational parameters same. In each cycle the mode is triggered at a different instant of time to construct the time evolution of mode frequency. The e-folding time of the charge evolution, suggesting a confinement time, has been obtained through a linear least square fit to a natural log of the measured frequencies. Multiple shots have been acquired at each instant to verify the shot-to-shot reproducibility and error in confinement time is obtained from goodness of fit.
\begin{figure}[ht!]
\begin{center}
\includegraphics[width=0.5\textwidth]{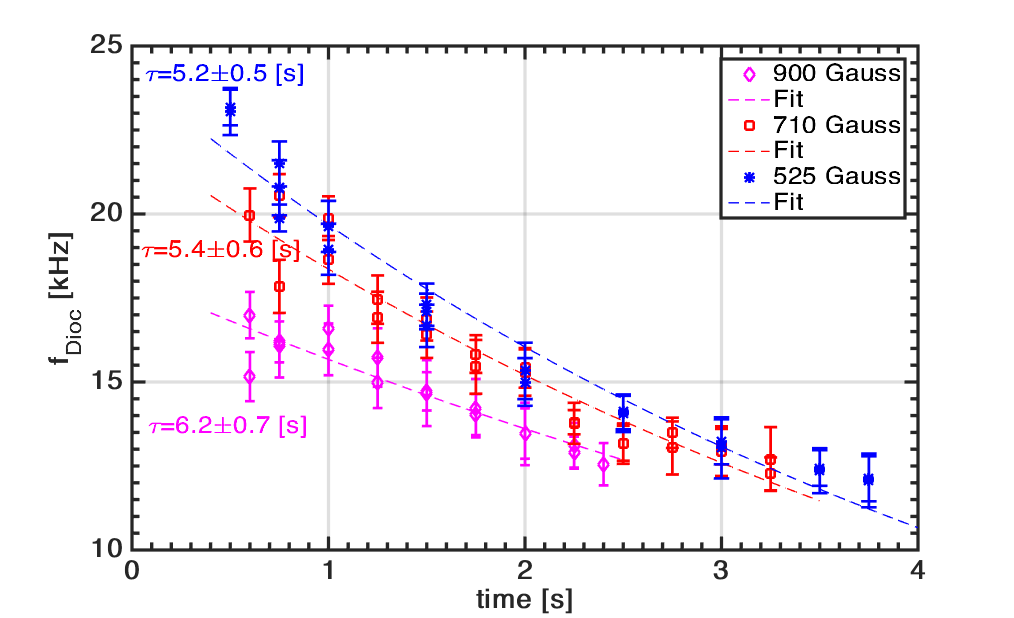}
\caption{Observed diocotron mode frequency on capacitive probe along with exponential fit for different magnetic fields at $1.5 \pm 0.5 \times 10^{-8}$ mbar pressure and injection energy $V_{Inj} = 250$ V.}
\label{Figure:Tau_1KGauss}
\end{center}
\end{figure}
Earlier experiments in SMARTEX-C had reported \cite{lavkesh_PoP_confinement_1S_2016} a confinement time of $\num{2.14} \pm \SI{0.1}{\second}$ at $380$ Gauss. The confinement seemed to be severly limited by droop $\sim 35\%$ in the B field. Pressures in the trapping region (following calibration) were of the order of $\num{1.5} \pm \SI{0.5e-8}{\milli\bar}$. Experiments with an upgraded TF coil (droop $<5\%$) were carried out at B fields between $\num{525}$ Gauss and $\num{900}$ Gauss (Fig. ~\ref{Figure:Tau_1KGauss}). The difference in initial frequencies at different B fields in Fig.~\ref{Figure:Tau_1KGauss} merits discussion. Soon after injection, the frequency of the 
spontaneously triggered diocotron mode is observed to follow the $1/B$ scaling, suggesting that initially injected charge is nearly same for all B fields. However, the damping of the mode and the rearrangement of the cloud that follows injection seem to be dependent on 
B field, involving different time scales. The charge left in the quiescent plasma is also different for different B fields. Admittedly, the dynamics and charge loss during the rearrangement is presently not well understood. However, after the initial dynamics, 
as the plasma becomes stable and quiescent, the loss of charge appears exponential in nature. Also, the confinement time, although showed marginal improvement ($\sim \num{5.2}-\SI{6.2}{\second}$), was interestingly, independent of B field. Currents were therefore lowered and experiments were conducted at reduced magnetic field of $\num{100}$ and $\num{200}$ Gauss. The droop in B field could therefore be further reduced to $< \num{0.1}\%$. More importantly, due to less Joule heating of TF coils, the duration of B field could be increased to $\SI{40}{\second}$. Further, to push the limits of confinement, pressures in the trapping region were lowered by a factor of $\sim 4$ ($\num{5.0} \pm \SI{1.0e-9}{\milli\bar}$) by turning on the NEG pump. The significant improvement in confinement is visible from Fig. ~\ref{Figure:Tau_100_200Gauss_100S}. At $\num{100}$ Gauss the confinement time is $\num{109.4} \pm \SI{9.3}{\second}$ and at $\num{200}$ Gauss it is $\num{102.8} \pm \SI{18.9}{\second}$, with goodness of fit $\sim \num{95}\%$. In addition to this being the highest confinement time with a purely toroidal B field, the independence of the lifetime with respect to B is being reported for the first time.   
The independence of B is a distinguishing characteristic of the Magnetic Pumping Transport theory. MPT is thought to fundamentally limit the confinement in toroidal traps due to the presence of an in-homogeneous B field, causing a gradual radial expansion of the plasma. Confinement time $\tau$ following dimensional analysis is given as \cite{stoneking_experimental_2009}
\begin{equation}
\tau = \frac{an_{\mathrm{av}}}{2\Gamma_{r}} 
\label{Eq:MPT_Arb}
\end{equation}
where $\Gamma_{r}$ is the flux rate derived by Crooks and O'Neil \cite{ONeil_Crooks_MagneticPumping_PoP1996}, 
\begin{equation}
\Gamma_{r} = \frac{1}{2}\nu_{\parallel,\perp}n(r)\frac{T_{\rm{eq}}}{-e\partial{\Phi}/{\partial r}}\left(\frac{r}{R_0}\right)^2
\label{Eq:Flux_Rate}
\end{equation}
where $n_{\mathrm{av}}$ is volume averaged electron density, $\partial{\Phi}/{\partial r} \sim 4\pi nea$, $T_{\rm{eq}}= (T_{\parallel}+2T_{\perp})/3$ and $\nu_{\parallel,\perp}$ is the rate at which electron-electron collisions equilibrate perpendicular and parallel temperatures. If one assumes $\nu_{\parallel,\perp}$ to be given by Spitzer rate of energy equipartition $\nu_{ee}$ ~\cite{RPP_V1_Leo_Brag}, $n_{\mathrm{av}} \sim n$, and $T_{\parallel} \approx T_{\perp}=T$, then Eq. (\ref{Eq:MPT_Arb}) becomes 
\begin{equation}
\tau_1 \simeq \frac{0.623}{\mathrm{ln}\Lambda_1} R_0^{2} \sqrt{T}
\label{Stoneking_tau}
\end{equation}
where $\mathrm{ln}\Lambda_1 = \mathrm{log}(\lambda_D/b)$ is Coulomb logarithm, $\lambda_D$ is Debye length, and $b = e^2/kT$ is the
distance of closest approach. If we use Eq. (\ref{Stoneking_tau}) the estimate of temperature would be $\sim 475$ eV for a plasma confinement of $100$ s ($R_{0}=\SI{13.5}{\centi\meter}$). Well-confined pure electron plasmas are typically assumed to be $1$ to few tens of $\si{\electronvolt}$. Even if we were to consider $T$ of few tens of eV, the unprecedented confinement time at lower pressures breaches the theoretical limits estimated from Eq. (\ref{Stoneking_tau}), fairly convincingly. This discrepancy thus warrants a closer scrutiny of the radial particle flux used to derive Eq. (\ref{Stoneking_tau}). It is possible that $\nu_{\parallel\perp}$ is not equal to $\nu_{ee}$, especially for plasmas with temperature anisotropy ($T_{\parallel} \ne T_{\perp}$). If instead, we use for $\nu_{\parallel\perp}$, the prediction given by modified Ichimaru and Rosenbluth formula \cite{Ichimaru_Rosenbluth_PoF_13_1970,Glinsky_PFB_1992} and confirmed experimentally ~\cite{Hyat_PRL_59.2975,Beck_PRL_68.317}, then the revised confinement time is given by,
\begin{equation}
\tau_2 \simeq \frac{2.2}{\mathrm{ln}\Lambda_2} R_0^{2} \sqrt{T}
\label{IR_tau}
\end{equation}
where $\mathrm{ln}\Lambda_2 = \left[\mathrm{log}(r_c/b) + 0.75\right]$, $r_c$ is thermal cyclotron radius, and this yields $T = 17$ eV. 
Note that, for weak magnetization ($r_c >> b$), the magnetic field dependence of the transport is expected to be very weak, as also seen from our experiment. It is thus shown that $100$ s confinement can be explained within the present framework of MPT considering the temperature anisotropy related collisional equiliberation rates.

In conclusion, recent experiments in SMARTEX-C have led us to confine pure electron plasmas with a purely toroidal magnetic field for 
approximately $\SI{100}{\second}$. Control of instabilities followed by improved operating parameters have led to such unprecedented 
confinement time. The lifetime appears to be independent of B field, suggesting magnetic pumping like transport. Availability of accurate temperature measurement will lead to better understanding of underlying transport mechanism.

Acknowledgments: Authors would like to thank Prof. Harishankar Ramachandran for useful comments made on data analysis, Ms. Minsha Shah for providing technical(electronics) support and Mr. Manu Bajpai for his support.

\bibliographystyle{apsrev}

\bibliography{NNP_PRL}
\end{document}